\documentclass[twoside]{article}
\usepackage{qic,epsfig,graphicx}
\usepackage{amsmath,amssymb}

\newcommand\eref[1]{Eq.~(\ref{eq:#1})}				  

\begin{document}
\setlength{\textheight}{8.0truein}    

\runninghead{Multiplayer quantum Minority game with decoherence}
            {Flitney and Hollenberg}

\normalsize\textlineskip
\thispagestyle{empty}
\setcounter{page}{1}

\copyrightheading{0}{0}{2005}{000--000}

\vspace*{0.88truein}

\alphfootnote

\fpage{1}

\centerline{\bf MULTIPLAYER QUANTUM MINORITY GAME WITH DECOHERENCE}
\vspace*{0.37truein}
\centerline{\footnotesize ADRIAN P. FLITNEY}
\vspace*{0.015truein}
\centerline{\footnotesize\it Centre for Quantum Computer Technology, School of Physics, University of Melbourne}
\baselineskip=10pt
\centerline{\footnotesize\it Parkville, VIC 3010, Australia}
\vspace*{10pt}
\centerline{\footnotesize LLOYD C. L. HOLLENBERG}
\vspace*{0.015truein}
\centerline{\footnotesize\it Centre for Quantum Computer Technology, School of Physics, University of Melbourne}
\baselineskip=10pt
\centerline{\footnotesize\it Parkville, VIC 3010, Australia}
\vspace*{0.225truein}
\publisher{(received date)}{(revised date)}

\vspace*{0.21truein}

\abstracts{
A quantum version of the Minority game for an arbitrary number of agents is considered.
It is known that when the number of agents is odd,
quantizing the game produces no advantage to the players,
but for an even number of agents new Nash equilibria appear
that have no classical analogue
and have improved payoffs.
We study the effect on the Nash equilibrium payoff
of various forms of decoherence.
As the number of players increases
the multipartite GHZ state becomes increasingly fragile,
as indicated by the smaller error probability required to reduce the
Nash equilibrium payoff to the classical level.
}{}{}

\vspace*{10pt}

\keywords{Quantum games, decoherence, Minority game, multiplayer games}
\vspace*{3pt}
\communicate{to be filled by the Editorial}

\vspace*{1pt}\textlineskip    

\section{INTRODUCTION}
\label{sec:intro}
Game theory is the formal description of conflict or competition situations
where the outcome is contingent upon the interaction
of the strategies of the various agents.
For every outcome, each player assigns a numerical measure
of the desirability to them of that outcome,
known as their utility or payoff.
(Strictly, the utility is a numerical measure
and the payoff is a relative ordering,
but for the purpose of the present work the two terms shall be used interchangeably.)
A solution of a game-theoretic problem
is a strategy profile that represents some form of equilibrium,
the best known of which is the Nash equilibrium (NE)~\cite{nash50}
from which no player can improve their payoff by a unilateral change in strategy.
Originally developed for use in economics~\cite{neumann44},
game theory is now a mature branch of mathematics used
in the social and biological sciences, computing and,
more recently, in the physical sciences~\cite{abbott02}.

The Minority game (MG),
initially proposed by Challet and Zhang~\cite{challet97},
has received much attention
as a model of a population of agents
repeatedly buying and selling in a market~\cite{savit99,johnson98,challet00,moro04}.
In its simplest form,
at each step the agents must independently select among a pair of choices,
labeled `0' and `1.'
Players selecting the least popular choice are rewarded with a unit payoff
while the majority emerge empty handed.
Players' strategies can be based on knowledge of previous selections
and successes in past rounds.
Examples of Minority games occur frequently in everyday life:
selecting a route to drive into the city,
choosing a checkout queue in the supermarket etc.
The idea behind the Minority game is neatly encapsulated by the following quote:
\begin{quote}
It is not worth an intelligent man's time to be in the majority.
By definition there are already enough people to do that---{\em Geoffery Harold Hardy}.
\end{quote}
The Minority game is generally restricted to an odd number of agents,
but even numbers can be permitted
with the proviso that when the number of players selecting 0 and 1 are equal
all players lose.

A game can be considered an information processing system,
where the players' strategies are the input
and the payoffs are the output.
With the advent of quantum computing
and the increasing interest in quantum information~\cite{nielsen00,eisert04}
it is natural to consider the combination of quantum mechanics
and game theory.
Papers by Meyer~\cite{meyer99} and Eisert {\em et al.}~\cite{eisert99}
paved the way for the creation of the new field of quantum game theory.
Classical probabilities are replaced by quantum amplitudes
and players can utilize superposition, entanglement and interference.

In quantum game theory, new ideas arise in two-player
\cite{eisert00,marinatto00,iqbal01a,iqbal02e,du02c,flitney02a,hr04}
and multiplayer settings~\cite{benjamin01b,kay01,du02b,han02a,iqbal02a,flitney04a,chen04}.
In the protocol of Eisert {\em et al.}~\cite{eisert99},
in two player quantum games there is no NE
when both players have access to the full set of unitary strategies~\cite{benjamin01a}.
Nash equilibria exist
amongst mixed quantum strategies~\cite{eisert00}
or when the strategy set is restricted in some way~\cite{eisert99,du01b,du03b}.
Strategies are referred to as pure
when the actions of the player at any stage is deterministic
and mixed when a randomizing device is used to select among actions.
That is, a mixed strategy is a convex linear combination of pure strategies.
In multiplayer quantum games
new NE amongst unitary strategies can arise~\cite{benjamin01b}.
These new equilibria have no classical analogues.
Reviews of quantum games and their applications are given by
Flitney and Abbott~\cite{flitney02c}
and Piotrowski and S{\l}adowski~\cite{piotr03c,piotr04a}.

The realization of quantum computing is still an endeavour
that faces great challenges~\cite{abbott03}.
A major hurdle is the maintenance of coherence during the computation,
without which the special features of quantum computation are lost.
Decoherence results
from the coupling of the system with the environment
and produces non-unitary dynamics.
Interaction with the environment can never be entirely eliminated
in any realistic quantum computer.
Zurek gives a review of the standard mechanisms of quantum decoherence~\cite{zurek03}. 
By encoding the logical qubits in a number of physical qubits,
quantum computing in the presence of noise is possible.
Quantum error correcting codes~\cite{preskill98}
function well provided the error rate is small enough,
while decoherence free subspaces~\cite{lidar03}
eliminate certain types of decoherence.

The theory of quantum control in the presence of noise is little studied.
Johnson has considered a three-player quantum game
where the initial state is flipped to $|111\rangle$ from the usual $|000\rangle$
with some probability~\cite{johnson01},
while \"{O}zdemir {\em et al.}~\cite{ozdemir04} have considered various two-player,
two strategy ($2 \times 2$) quantum games
where the initial state is corrupted by bit flip errors.
In both papers it was found that quantum effects impede the players:
above a certain level of noise
they are then
better off playing the classical game.
Chen {\em et al.} found that the NE in a set of restricted quantum strategies
was unaffected by decoherence
in quantum Prisoners' Dilemma~\cite{chen03a},
while Jing-Ling Chen {\em et al.} have considered Meyer's quantum Penny Flip
game with various forms of decoherence~\cite{chen02b}.
Decoherence in various two player quantum games
in the Eisert protocol is considered by Flitney and Abbott~\cite{flitney04d,flitney05}.
The quantum player maintains an advantage over a player restricted to classical strategies
provided some level of coherence remains.
The current work reviews the formalism for quantum games with decoherence
and discusses the existing results for the quantum Minority game,
before considering the quantum Minority game in the presence of decoherence.

\section{QUANTUM GAMES WITH DECOHERENCE}
\label{sec:games}
The standard protocol for quantizing a game
is well described in a number of papers~\cite{eisert99,flitney02c,lee03b}
and will be covered here only briefly.
If an agent has a choice between two strategies,
the selection can be encoded in the classical case by a bit.
To translate this into the quantum realm the bit is altered to a qubit,
with the computational basis states $|0\rangle$ and $|1\rangle$ representing
the original classical strategies.
The initial game state consists of one qubit for each player,
prepared in an entangled GHZ state
by an entangling operator $\hat{J}$ acting on $|00 \ldots 0\rangle$.
Pure quantum strategies are local unitary operators acting on a player's qubit.
After all players have executed their moves the game state undergoes
a positive operator valued measurement
and the payoffs are determined from the classical payoff matrix.
In the Eisert protocol this is achieved by applying $\hat{J}^{\dagger}$
to the game state and then making a measurement in the computational basis state.
That is, the state prior to the measurement in the $N$-player case can be computed by
\begin{eqnarray}
\label{eq:qgame}
|\psi_0\rangle &=& |00 \ldots 0\rangle \\ \nonumber
|\psi_1\rangle &=& \hat{J} |\psi_0\rangle \\ \nonumber
|\psi_2\rangle &=& (\hat{M}_1 \otimes \hat{M}_2 \otimes \ldots \otimes \hat{M}_N)  |\psi_1\rangle \\ \nonumber
|\psi_f\rangle &=& \hat{J}^{\dagger} |\psi_2\rangle,
\end{eqnarray}
where
$|\psi_0 \rangle$ is the initial state of the $N$ qubits,
and $\hat{M}_k, \; k = 1,\ldots,N$, is a unitary operator representing the move of player $k$.
The classical pure strategies
are represented by the identity and the bit flip operator.
The entangling operator $\hat{J}$ commutes with any direct product of classical moves,
so the classical game is simply reproduced if all players select a classical move.

Lee and Johnson~\cite{lee03b} describe a more general quantum game protocol
where the prepared initial state need not be a GHZ state
and the final measurement need not be in the computational basis.
Their protocol includes the method of Eisert.
Wu~\cite{wu05} considers a further generalization to a game on quantum objects.

To consider decoherence it is most convenient to use the density matrix notation
for the state of the system
and the operator sum representation for the quantum operators.
There are known limitations of this representation~\cite{royer96};
a variety of other techniques for calculating decoherence
are considered by Brandt~\cite{brandt98}.
Decoherence includes dephasing,
which randomizes the relative phase between the $|0\rangle$ and $|1\rangle$ states,
and dissipation, that modifies the populations of the states,
amongst other forms~\cite{nielsen00}.
Pure dephasing can be expressed at the state level as
\begin{equation}
        a |0\rangle \:+\: b |1\rangle
                \rightarrow a |0\rangle \:+\: b \, e^{i \phi} |1\rangle.
\end{equation}
If the phase shift $\phi$ is a random variable
with a Gaussian distribution of mean zero and variance $2 \lambda$,
the density matrix obtained after averaging over all values of $\phi$ is~\cite{nielsen00}
\begin{equation}
	\left( \begin{array}{cc}
                |a|^2 & a \bar{b} \\
                \bar{a} b & |b|^2
	\end{array} \right)
        \rightarrow \left( \begin{array}{cc}
                |a|^2 & a \bar{b} \, e^{-\lambda} \\
                \bar{a} b \, e^{-\lambda} & |b|^2
	\end{array} \right).
\end{equation}
Thus, over time, dephasing causes an exponential decay
of the off-diagonal elements of the density matrix
and so is also know as phase damping.

An example of dissipation is amplitude damping.
This could correspond, for example, to loss of a photon in an optical system.
The effect on the density matrix is to reduce
the amplitude of $|1\rangle \langle 1|$ as well as the off-diagonal elements.

Making a measurement with probability $p$
in the $\{|0\rangle, |1\rangle \}$ basis
on a qubit described by the density matrix $\rho$
can be represented in the operator sum formalism by
\begin{equation}
\rho \; \rightarrow\; \sum_{j=0}^{2} \hat{{\cal E}}_j \, \rho \, \hat{{\cal E}}_j^{\dagger},
\end{equation}
where ${\cal E}_0 = \sqrt{p} \, |0\rangle \langle 0|$,
$\hat{{\cal E}}_1 = \sqrt{p}\,  |1\rangle \langle 1|$,
and $\hat{{\cal E}}_2 = \sqrt{1-p} \, \hat{I}$.
By the addition of further $\hat{{\cal E}}_j$'s an extension to $N$ qubits is achieved:
\begin{equation}
\label{eq:measure}
\rho \; \rightarrow\; \sum_{j_1, \ldots, j_N = 0}^{2}
	\hat{{\cal E}}_{j_1} \otimes \ldots \otimes \hat{{\cal E}}_{j_N} \, \rho \,
		(\hat{{\cal E}}_{j_1} \otimes \ldots \otimes \hat{{\cal E}}_{j_N})^{\dagger},
\end{equation}
where, here, $\rho$ is an $N$-qubit state.
By identifying $1-p = e^{-\lambda}$,
the measurement process as described has the same results as pure dephasing:
the exponential decay of the off-diagonal elements of $\rho$.

In quantum computing it is usual to consider single qubit errors caused by bit flips,
$\rho \rightarrow \hat{\sigma}_x \, \rho \, \hat{\sigma}_x$,
phase flips, $\rho \rightarrow \hat{\sigma}_z \, \rho \, \hat{\sigma}_z$,
and bit-phase flips, $\rho \rightarrow \hat{\sigma}_y \, \rho \, \hat{\sigma}_y$.
Depolarization is a process where by a quantum state decays to an equal mixture
of the $|0\rangle$ and $|1\rangle$ states.
It can be considered to be
a combination of bit, phase, and bit-phase flip errors:
\begin{eqnarray}
\label{eq:depolar}
\rho &\rightarrow& \frac{p I}{2} \:+\: (1-p) \, \rho \\ \nonumber
     &=& (1 - p) \, \rho \:+\: \frac{p}{3} \, \hat{\sigma}_x \, \rho \, \hat{\sigma}_x +
						\hat{\sigma}_y \, \rho \, \hat{\sigma}_y +
						\hat{\sigma}_z \, \rho \, \hat{\sigma}_z,
\end{eqnarray}
where $I/2 = (|0\rangle \langle 0| + |1\rangle \langle 1|)/2$ is the completely mixed state.
The errors given above are by no means an exhaustive list
but consideration of them will give a good indication of the behaviour
of our quantum system subject to random decoherence.

The physical implementation of a quantum system determines when the decoherence operators
should be inserted in the formalism.
For example,
in a solid state implementation, errors,
including qubit memory errors, need to be considered after each time step,
while in an optical implementation memory errors only arise from infrequent photon loss,
but errors need to be associated with each quantum gate.
In addition, there may be errors occurring in the final measurement process.
In this paper we shall describe
a quantum game in the Eisert scheme with decoherence in the following manner
\begin{eqnarray}
\label{eq:scheme}
	\rho_0 =& | \psi_0 \rangle \langle \psi_0 |
			& {\rm (initial \; state)} \\ \nonumber
	\rho_1 =& \hat{J} \rho_0 \hat{J}^{\dagger} & \mbox{(entanglement)} \\ \nonumber
	\rho_2 =& D(\rho_1, p) & \mbox{(partial decoherence)} \\ \nonumber
	\rho_3 =& (\otimes_{k=1}^{N} \hat{M}_{k}) \, \rho_2 \,
			(\otimes_{k=1}^{N} \hat{M}_{k})^{\dagger} \qquad
		& \mbox{(players' moves)} \\ \nonumber
	\rho_4 =& D(\rho_3, p')
		& \mbox{(partial decoherence)} \\ \nonumber
	\rho_5 =& \hat{J}^{\dagger} \rho_4 \hat{J} &
		\mbox{(preparation for measurement)},
\end{eqnarray}
to produce the final state $\rho_f \equiv \rho_5$ upon which a measurement is taken.
That is,
errors are considered after the initial entanglement
and after the players' moves.
In all subsequent calculations we set $p'=p$.
An additional error possibility could be included after the $\hat{J}^{\dagger}$ gate
but this gate is not relevant in the quantum Minority game
since it only mixes states with the same player(s) winning~\cite{benjamin01b}.
Hence the gate and any associated decoherence will be omitted for the remainder of the paper.
The $\hat{J}$ gate can be implemented by a (generalized) Hadamard gate
followed by a sequence of CNOT gates,
as indicated in figure~\ref{f:J}.
When the number of qubits is large
the possibility of errors occurring within the $J$ gate needs to be considered
but is not done so here.
The function $D(\rho, p)$ is a completely positive map
that applies some form of decoherence to the state $\rho$
controlled by the probability $p$.
For example, for bit flip errors
\begin{equation}
D(\rho, p) = (\sqrt{p} \, \hat{\sigma}_x + \sqrt{1-p} \, \hat{I})^{\otimes N} \rho \:
		 (\sqrt{p} \, \hat{\sigma}_x + \sqrt{1-p} \, \hat{I})^{\otimes N}.
\end{equation}
 
The scheme of \eref{scheme} is shown in figure \ref{f:qgame}.
The expectation value of the payoff to the $k$th player is
\begin{equation}
	\langle \$^k \rangle =
		\sum_{\xi} \hat{\cal P}_{\xi} \, \rho_f
			\, \hat{\cal P}_{\xi}^{\dagger} \, \$_{\xi}^{k},
\end{equation}
where $\hat{\cal P}_{\xi} = |\xi \rangle \langle \xi|$
is the projector onto the computational state $|\xi \rangle$,
$\$_{\xi}^{k}$ is the payoff to the $k$th player
when the final state is $|\xi\rangle$,
and the summation is taken over 
$\xi = j_1 j_2 \ldots j_N, \; j_i \in \{0,1\}$.

\begin{figure}
\begin{center}
\begin{picture}(160,100)(0,0)
	\multiput(0,20)(0,30){3}{$|0\rangle$}
	\multiput(20,25)(0,30){2}{\line(1,0){160}}
	\put(20,85){\line(1,0){20}}
	\put(40,80){\framebox(12,12){$\hat{H}$}}
	\put(52,85){\line(1,0){128}}
	\multiput(80,85)(20,-30){3}{\circle*{6}}
	\multiput(80,85)(20,-30){2}{\line(0,-1){32}}
	\put(120,25){\line(0,-1){20}}
	\multiput(80,55)(20,-30){2}{\circle{6}}
	\multiput(10,0)(90,0){2}{$\vdots$}
\end{picture}
\end{center}
\fcaption{\label{f:J}A possible gate sequence to implement the entangling $\hat{J}$ gate.}
\end{figure}

\begin{figure}
\begin{center}
\begin{picture}(260,160)(0,-30)
        \multiput(0,58)(0,30){2}{$|0\rangle$}
        \put(0,8){$|0\rangle$}
        \put(0,73){$\otimes$}
        \put(240,45){$|\psi_f\rangle$}

        \multiput(20,60)(40,0){6}{\line(1,0){20}}
        \multiput(20,90)(40,0){6}{\line(1,0){20}}
        \put(45,70){$\hat{J}$}
        \put(205,70){$\hat{J}^{\dagger}$}

	\multiput(40,50)(20,0){2}{\line(0,1){50}}
	\multiput(200,50)(20,0){2}{\line(0,1){50}}
	\multiput(40,100)(160,0){2}{\line(1,0){20}}

        \put(120,80){\framebox(20,20){$\hat{M}_1$}}
        \put(120,50){\framebox(20,20){$\hat{M}_2$}}
        \put(120,0){\framebox(20,20){$\hat{M}_N$}}
	\multiput(5,30)(125,0){2}{$\vdots$}

	\multiput(40,0)(160,0){2}{\line(1,0){20}}
	\multiput(40,0)(20,0){2}{\line(0,1){20}}
	\multiput(200,0)(20,0){2}{\line(0,1){20}}
        \multiput(20,10)(40,0){6}{\line(1,0){20}}
	
	\put(90,50){\oval(20,100)[t]}
	\put(90,20){\oval(20,40)[b]}
        \put(170,50){\oval(20,100)[t]}
        \put(170,20){\oval(20,40)[b]}
	\multiput(87,70)(80,0){2}{$D$}

	\multiput(40,25)(20,0){4}{\line(0,1){4}}
	\multiput(40,33)(20,0){4}{\line(0,1){4}}
	\multiput(40,41)(20,0){4}{\line(0,1){4}}
	\multiput(160,25)(20,0){4}{\line(0,1){4}}
	\multiput(160,33)(20,0){4}{\line(0,1){4}}
	\multiput(160,41)(20,0){4}{\line(0,1){4}}

	\put(90,110){\circle*{10}}
	\put(170,-10){\circle*{10}}

	\put(90,105){\line(0,-1){5}}
	\put(170,-5){\line(0,1){5}}
	\put(20,110){\line(1,0){65}}
	\put(20,-10){\line(1,0){145}}




	\put(1,107){$p$}
	\put(1,-13){$p'$}

        \put(110,120){\vector(1,0){60}}
        \put(125,125){time}

\end{picture}
\end{center}
\fcaption{\label{f:qgame}The flow of information in an $N$-person quantum game
with decoherence,
where $M_k$ is the move of the $k$th player
and $\hat{J}$ ($\hat{J}^{\dagger}$) is an entangling (dis-entangling) gate.
The central horizontal lines are the players' qubits and
the top and bottom lines are classical random bits with a probability
$p$ or $p'$, respectively, of being 1.
Here, $D$ is some form of decoherence
controlled by the classical bits.
Figure from Flitney and Abbott~\cite{flitney05}.}
\end{figure}


\section{RESULTS FOR THE MULTIPLAYER MINORITY GAME}
\label{sec-minority}
\subsection{Without decoherence}
In the classical Minority game the equilibrium is trivial:
a maximum expected payoff is achieved if all players base their decision on the toss of a fair coin.
The interest lies in studying the fluctuations that arise when agents
use knowledge of past behaviour to predict a successful option for the next play.
In the quantum game,
as we shall see,
a more efficient equilibrium can arise
when the number of players is even.
This paper only considers the situation where
players do not make use of their knowledge of past behaviour.
The classical pure strategies are then ``always choose 0'' or ``always choose 1.''
A pure quantum strategy is an SU(2) operator:
\begin{equation}
\label{e-qstrategy}
\hat{M}(\theta, \alpha, \beta) =
	\left( \begin{array}{cc}
		e^{i \alpha} \cos (\theta/2) & i e^{i \beta} \sin (\theta/2) \\
		i e^{-i \beta} \sin (\theta/2) & e^{-i \alpha} \cos(\theta/2)
	\end{array} \right),
\end{equation}
where $\theta \in [ 0,\pi ]$ and $\alpha, \beta \in [ -\pi,\pi ]$.
The $k$th player's move is $\hat{M}_k(\theta_k, \alpha_k, \beta_k)$.
Here, $\hat{I} \equiv \hat{M}(0,0,0)$ and $i \hat{X} \equiv \hat{M}(\pi,0,0) = i \hat{\sigma}_x$
correspond to the two classical pure strategies.
Entanglement, controlled by a parameter $\gamma \in [0, \pi/2]$, is achieved by
\begin{equation}
\label{eq:J}
\hat{J}(\gamma) = \exp \left( i \frac{\gamma}{2} \, \sigma_x^{\otimes N} \right),
\end{equation}
with $\gamma = \pi/2$ corresponding to maximal entanglement in a GHZ state.
Operators of the form $\hat{M}(\theta,0,0)$
are equivalent to classical mixed strategies, with the mixing controlled by $\theta$,
since when all players use these strategies the quantum game reduces to the classical one.
There is some arbitrariness about the representation of the operators.
Other representations may lead to a different overall phase in the final state
but this has no physical significance.

Benjamin and Hayden showed that in the four player quantum MG
an optimal strategy arises~\cite{benjamin01b}:
\begin{eqnarray}
\label{eq:ne4}
\hat{s}_{\rm \scriptscriptstyle NE}
			&=& \frac{1}{\sqrt{2}} \cos (\frac{\pi}{16})(\hat{I} + i \hat{\sigma}_x)
		 	 \:-\: \frac{1}{\sqrt{2}} \sin (\frac{\pi}{16})(i \hat{\sigma}_y + i \hat{\sigma}_z) \\ \nonumber
			&=& \hat{M}(\frac{\pi}{2}, \frac{-\pi}{16}, \frac{\pi}{16}).
\end{eqnarray}
The strategy profile
$\{ \hat{s}_{\rm \scriptscriptstyle NE}, \hat{s}_{\rm \scriptscriptstyle NE},
\hat{s}_{\rm \scriptscriptstyle NE}, \hat{s}_{\rm \scriptscriptstyle NE} \}$
results in a NE with an expected payoff of $\frac{1}{4}$
to each player,
the maximum possible from a symmetric strategy profile,
and twice that that can be achieved in the classical game,
where the players can do no better than selecting 0 or 1 at random.
The optimization is the result of the elimination
of the states for which no player scores:
those where all the players make the same selection
or where the choices are balanced.

\begin{figure}
\begin{center}
\includegraphics[width=8cm]{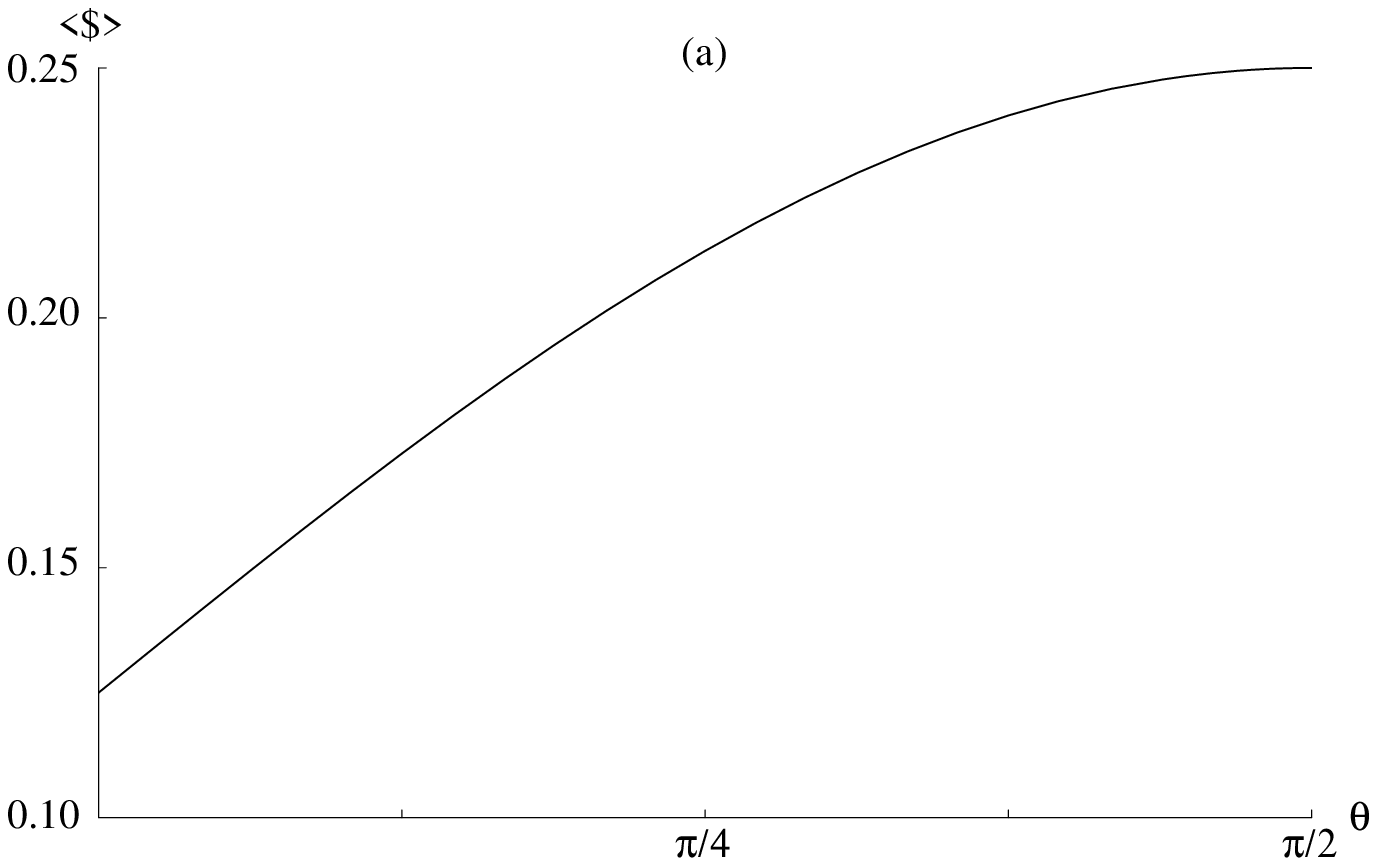}

\vspace*{1cm}

\includegraphics[width=8cm]{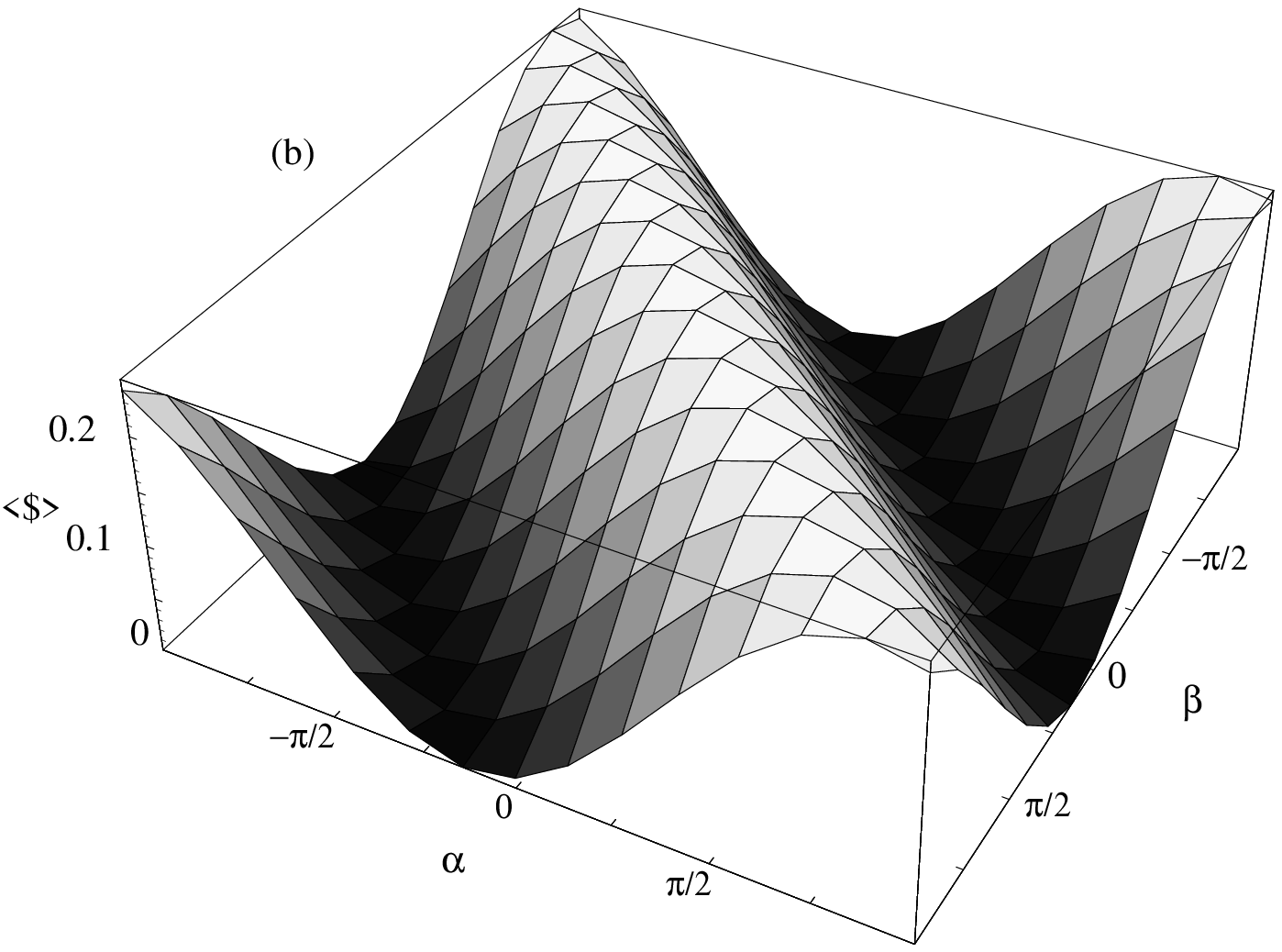}
\end{center}
\fcaption{\label{f:min4-ne}The payoff to the first player in an $N=4$ player quantum Minority game
without decoherence
when they choose the strategy (a) $\hat{M}(\theta, -\pi/16, \pi/16)$
or (b) $\hat{M}(\pi/2, \alpha, \beta)$,
while the other players all select $\hat{s}_{\rm \scriptscriptstyle NE}$.}
\end{figure}

The strategy of \eref{ne4} is seen to be a NE by observing
the payoff to the first player when they vary from the NE profile
by selecting the general strategy $\hat{M}(\theta, \alpha, \beta)$
while the others play $\hat{s}_{\rm \scriptscriptstyle NE}$.
Figure~\ref{f:min4-ne} shows the first player's payoff
as a function of $\theta$ when $\beta = -\alpha = \pi/16$,
and as a function of $\alpha$ and $\beta$ when $\theta = \pi/2$.
The latter figure indicates that the NE is not strict:
varying the strategy to
$\hat{M}(\pi/2,\, \eta-\pi/16,\, \eta+\pi/16)$,
for arbitrary $\eta \in \{ -15 \pi/16,\, 15 \pi/16 \}$
leaves the payoff unchanged.
Specifically, the payoff to the first player when they play $\hat{M}(\theta, \alpha, \beta)$
while the others select $\hat{s}_{\rm \scriptscriptstyle NE}$ is
\begin{equation}
\langle \$ \rangle = \frac{1}{8} + \frac{1}{8} \cos(\frac{\pi}{8} + \alpha - \beta)\, \sin \theta
\end{equation}
which is maximized when $\theta = \pi/2$ and $\pi/8 + \alpha - \beta = 2 n \pi$ for integer $n$.
This demonstrates that the strategy given in \eref{ne4} is a NE.

\begin{figure}
\begin{center}
\includegraphics[width=8cm]{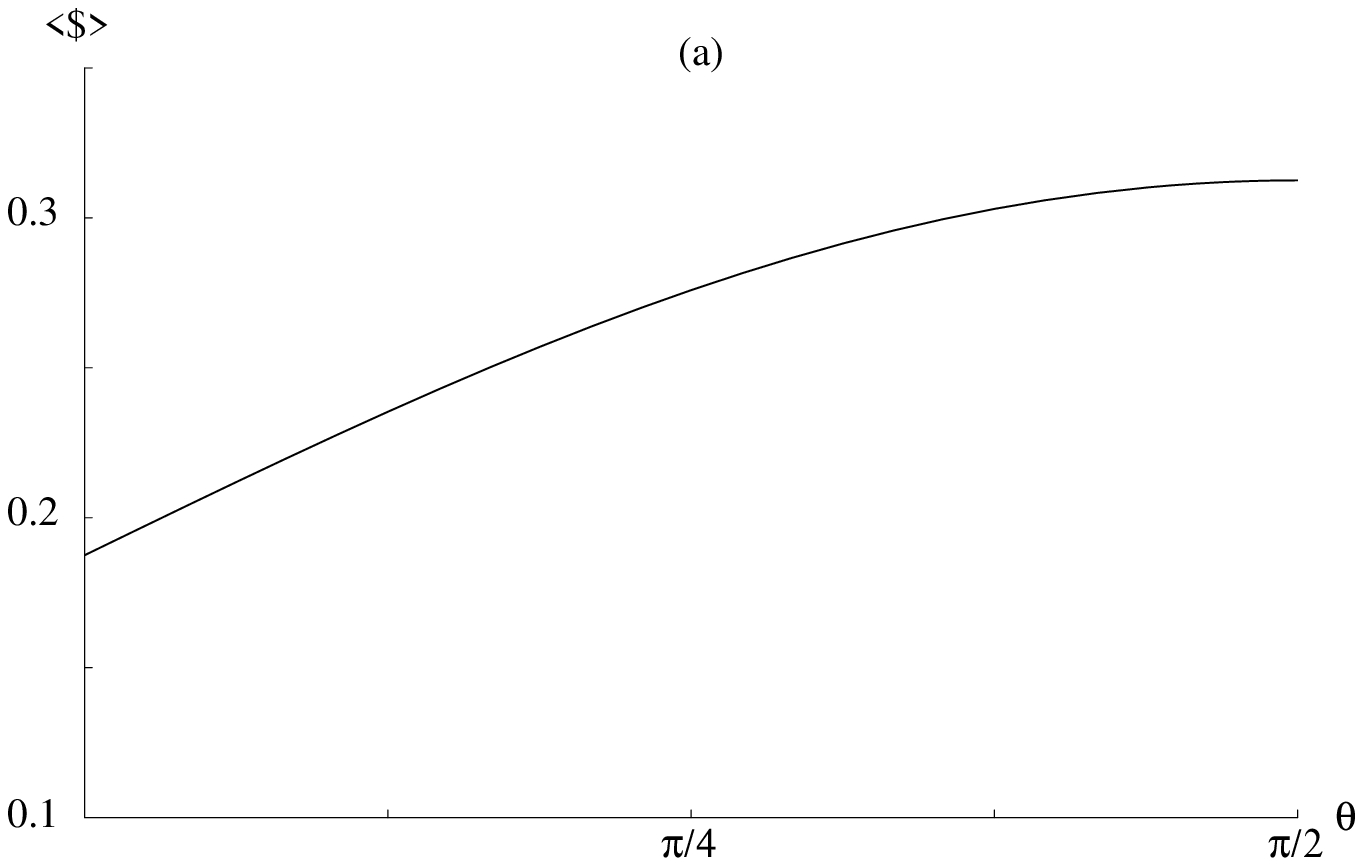}

\vspace*{1cm}

\includegraphics[width=8cm]{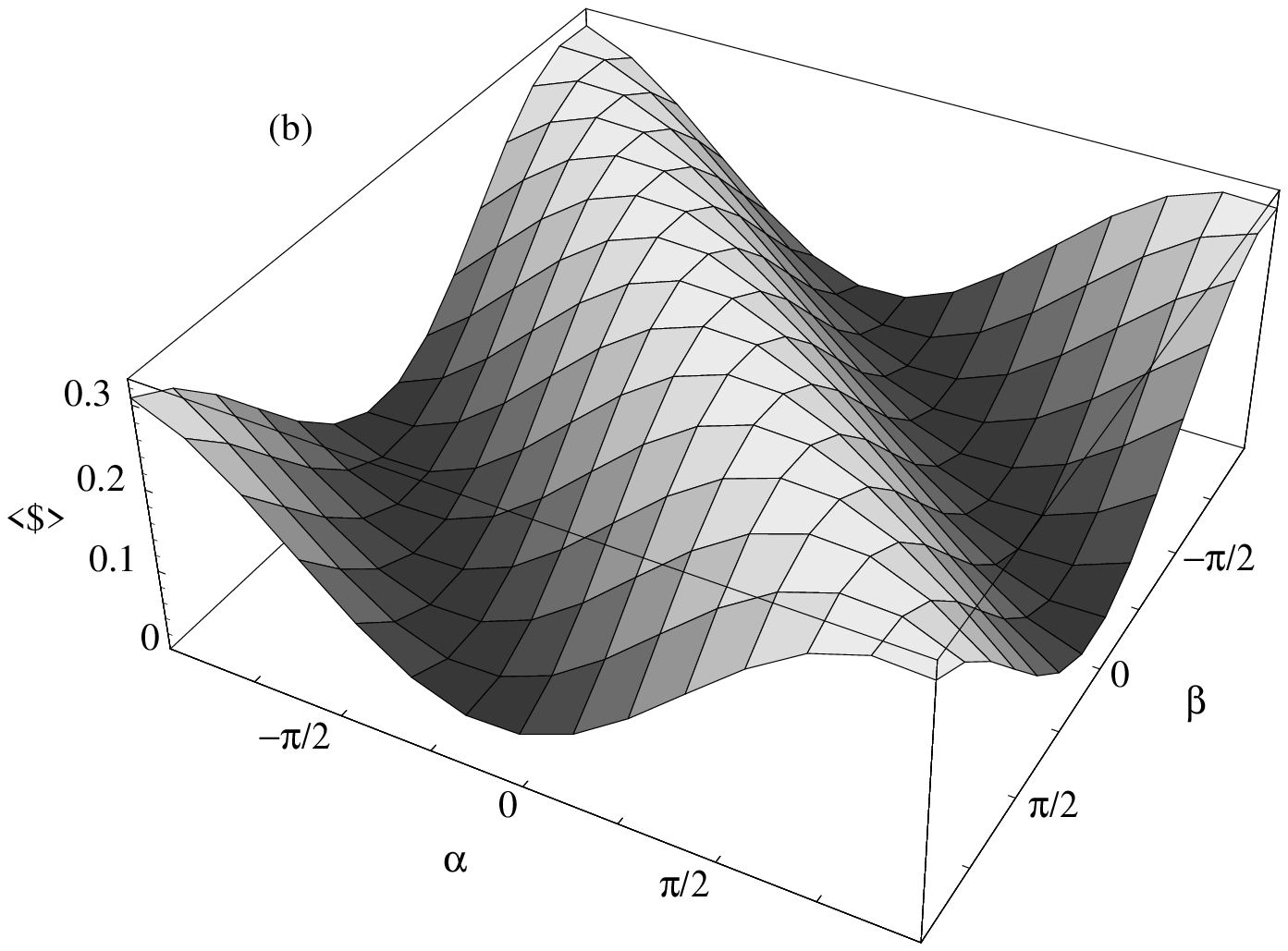}
\end{center}
\fcaption{\label{f:min6-ne}The payoff to the first player in an $N=6$ player quantum Minority game
without decoherence
when they choose the strategy (a) $\hat{M}(\theta, -\pi/24, \pi/24)$
or (b) $\hat{M}(\pi/2, \alpha, \beta)$,
while the other players all select $\hat{s}_{\rm \scriptscriptstyle NE}$.} 
\end{figure}

This result has been generalized to arbitrary $N$ by Chen {\it et al.}
who show that analogous NE occur
for all even $N$~\cite{chen04}.
A way of arriving at this result in our notation,
with $p=p'=0$,
is to consider a symmetric strategy profile where all players choose
\begin{equation}
\hat{s}_{\delta} = \hat{M}(\frac{\pi}{2}, -\delta, \delta),
\end{equation}
for some $\delta \in \mathbb{R}$ to be determined.
Then,
\begin{equation}
|\psi_f\rangle = \left[ \frac{1}{\sqrt{2}} \left(
			\begin{array}{cc}
				e^{-i \delta} & i e^{i \delta} \\
				i e^{-i \delta} & e^{i \delta}
			\end{array}
			\right) \right]^{\otimes N}
				(|00 \ldots 0\rangle + i |11 \ldots 1\rangle).
\end{equation}
Thus, for even $N$,
the coefficient of states in $|\psi_f\rangle$ that have an equal number of ones and zeros is proportional to
\begin{equation}
(e^{-i \delta})^{N/2} (i e^{-i \delta})^{N/2} + i (e^{i \delta})^{N/2} (i e^{i \delta})^{N/2}
	= \pm (1+i) (\cos (N \delta) - \sin (N \delta)),
\end{equation}
giving a probability for these states proportional to
$1 - \sin (2 N \delta)$.
This probability vanishes when
\begin{equation}
\delta = \frac{(4 n + 1) \pi}{4 N}, \qquad n = 0, \pm 1, \pm 2, \ldots
\end{equation}
For the collective good,
the vanishing of the balanced states is optimal
since these are the ones for which no player scores.
Each value of $\delta$ gives a NE for the $N$ even player quantum MG.
In addition,
for each $\delta$ there is a continuum of symmetric NE strategies of the form
$\hat{M}(\pi/2, \eta - \delta, \eta + \delta)$.
Figures~\ref{f:min6-ne} indicates that $\hat{M}(-\pi/2, -\pi/24, \pi/24)$
is a NE for the six player MG.
For $N>4$ the payoffs for the NE strategies are not Pareto optimal.
For example, for $N=6$ each player scores $\frac{5}{16}$
compared with the Pareto optimal payoff of $\frac{1}{3}$
that would result if all the final states consisted of two players
selecting one option while the other four chose the second option.
An alternate way of expressing the lack of optimality
is to say that the final state prior to measurement,
in the six player game,
has a probability of $\frac{15}{16}$ of giving a Pareto optimal
result when a measurement of the state is taken in the computational basis.

The NE that arises from selecting $\delta = \pi/(4 N)$ and $\eta = 0$
may serve as a focal point for the players
and be selected in preference to the other equilibria.
However,
if the players select $\hat{s}_{\scriptscriptstyle NE}$
corresponding to different values of $n$ the result may not be a NE.
For example, in the four player MG,
if the players select $n_A, n_B, n_C$, and $n_D$, respectively,
the resulting payoff depends on $(n_A + n_B + n_C + n_D) ({\rm mod}\, 4)$.
If the value is zero, all players receive the quantum NE payoff of $\frac{1}{4}$,
if it is one or three, the expected payoff is reduced to the classical NE value of $\frac{1}{8}$,
while if it is two, the expected payoff vanishes.
As a result,
if all the players choose a random value of $n$
the expected payoff is the same as that for the classical game ($\frac{1}{8}$)
where all the players selected 0 or 1 with equal probability.
Analogous results hold for the quantum MG with larger numbers of players.

When $N$ is odd the situation is changed.
The Pareto optimal situation
would be for $(N-1)/2$ players to select one alternative
and the remainder to select the other.
In this way the number of players that receive a reward is maximized.
In the entangled quantum game there is no way to achieve this with a symmetric strategy profile.
Indeed, all quantum strategies reduce to classical ones
and the players can achieve no improvement in their expected payoffs~\cite{chen04}.

\begin{figure}
\begin{center}
\includegraphics[width=8cm]{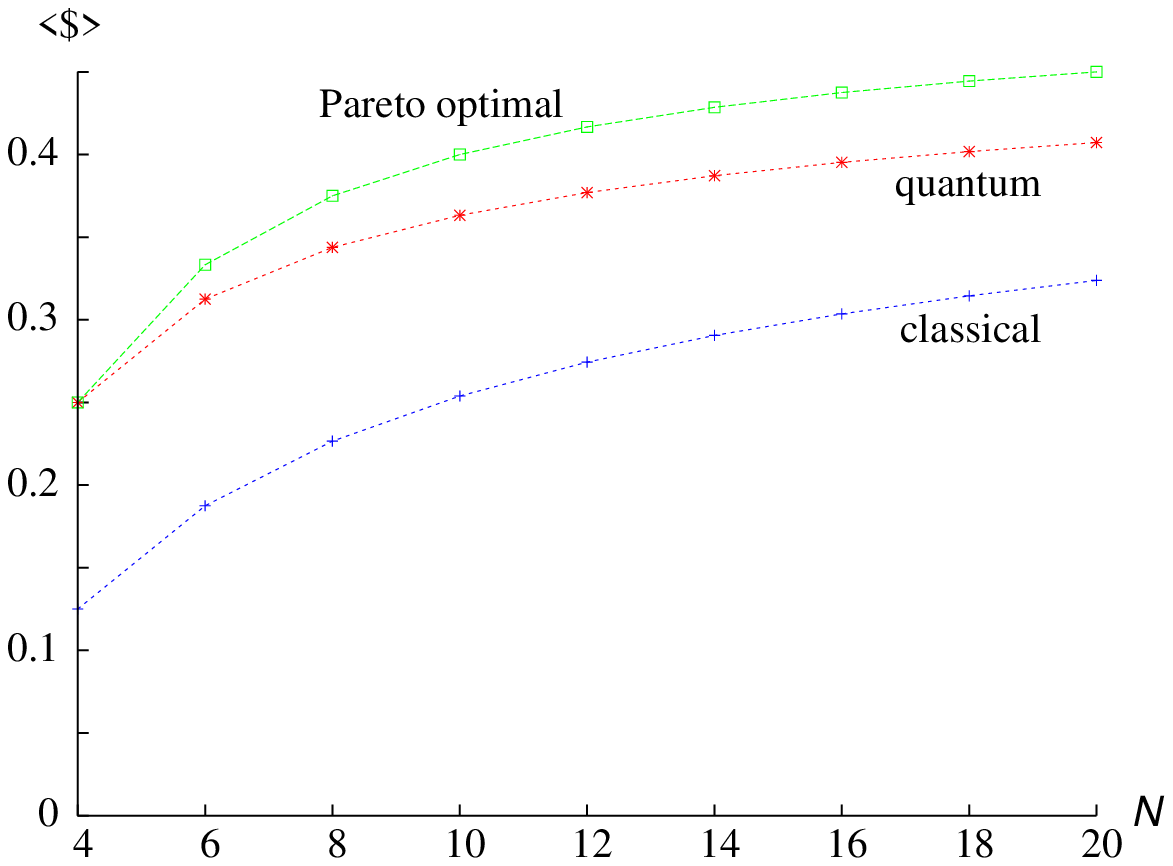}
\end{center}
\fcaption{\label{f:Npayoffs}The Nash equilibrium payoff as a function of the number of players ($N$)
for even $N$ for the fully entangled quantum case (*)
and the classical case (+).
Compare with the Pareto optimal payoffs ($\scriptstyle \Box$).
The curves slowly converge to $\langle \$ \rangle = \frac{1}{2}$ as $N \rightarrow \infty$.}
\end{figure}

The NE payoff for the $N$ even quantum game is precisely
that of the $N-1$ player classical game where each player selects 0 or 1 with equal probability.
The effect of the entanglement and the appropriate choice of strategy is
to eliminate some of the least desired final states,
those with equal numbers of zeros and ones.
The difference in behaviour between odd and even $N$ arises since,
although in both cases the players can arrange for the final state
to consist of a superposition with only even (or only odd) numbers of zeros,
only in the case when $N$ is even is this an advantage to the players.
Figure \ref{f:Npayoffs} shows the maximum expected payoffs
for the quantum and classical MG for even $N$.

\subsection{With decoherence}
\begin{figure}
\begin{center}
\includegraphics[width=9cm]{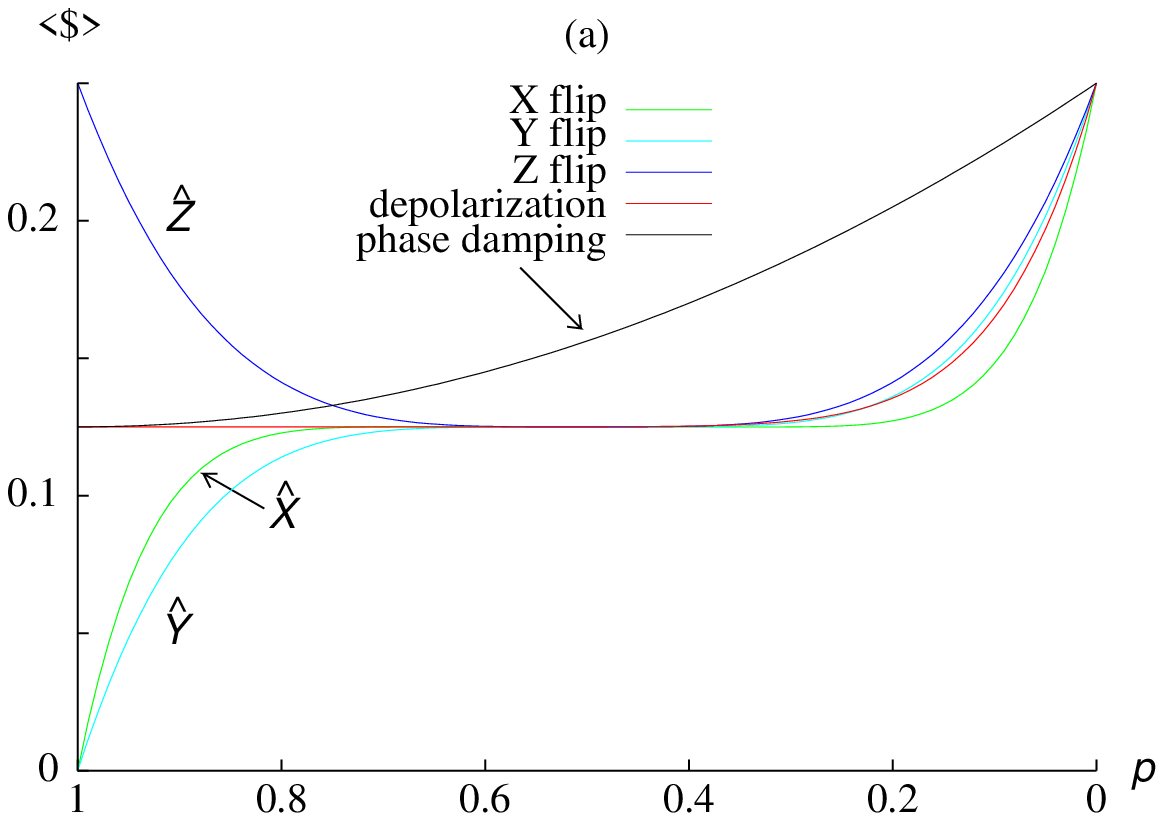}

\vspace*{1cm}

\includegraphics[width=9cm]{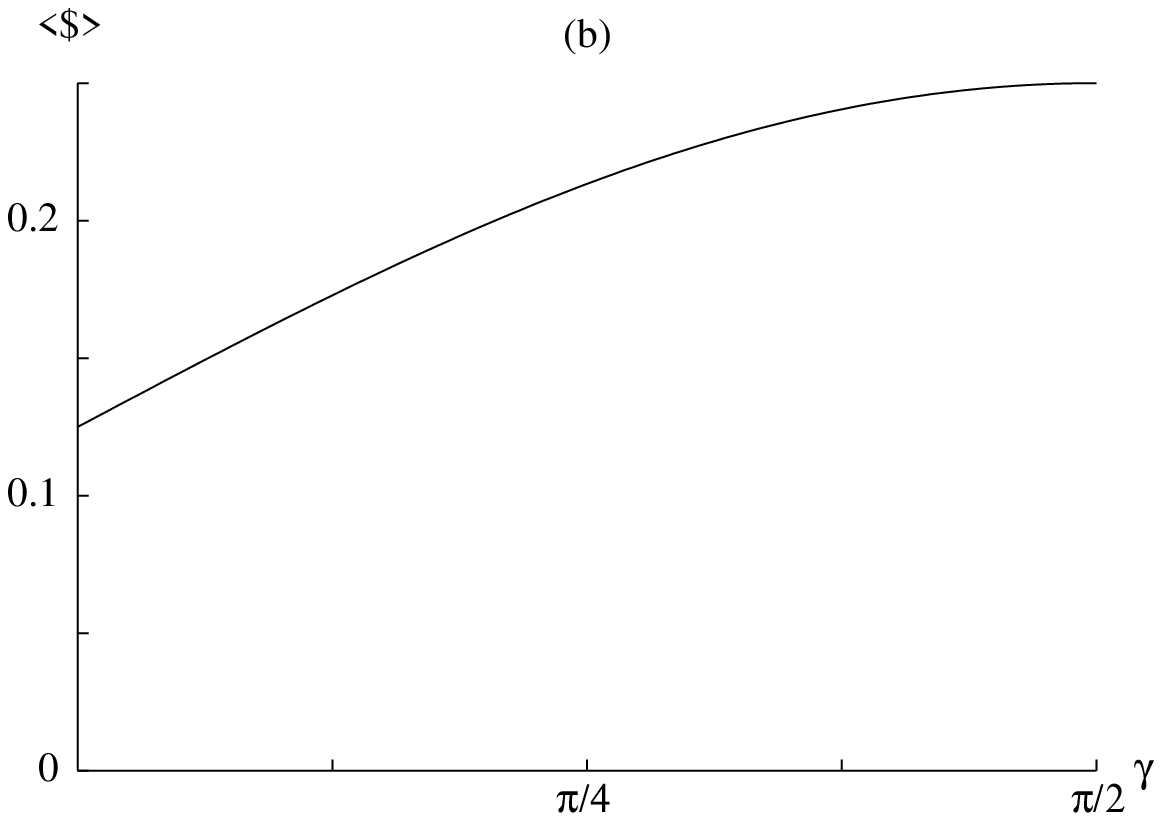}
\end{center}
\fcaption{\label{f:min4-decoh}(a) The Nash equilibrium payoff in an $N=4$ player quantum Minority game
as a function of the decoherence probability $p$.
The decoherence goes from the unperturbed quantum game at $p=0$ (right)
to maximum decoherence at $p=1$ (left).
The curves indicate decoherence by phase damping (black),
depolarization (red),
bit flip errors (green),
phase flip errors (blue)
and bit-phase flip errors (blue-green).
Compare this with (b) the Nash equilibrium payoff for $N=4$ as a function of the entangling
parameter $\gamma$ [\eref{J}].}
\end{figure}

The addition of decoherence by dephasing (or measurement)
to the four player quantum MG
results in a gradual diminution of the NE payoff,
ultimately to the classical value of $\frac{1}{8}$
when the decoherence probability $p$ is maximized,
as indicated in figure~\ref{f:min4-decoh}.
However, the strategy given by \eref{ne4} remains a NE for all $p < 1$.
This is in contrast with the results of Johnson~\cite{johnson01}
for the three player ``El Farol bar problem''~\cite{arthur94}
and \"{O}zdemir {\em et al.}~\cite{ozdemir04}
for various two player games in the Eisert scheme,
who showed that the quantum optimization
did not survive above a certain noise threshold
in the quantum games they considered.
Bit (or $\hat{X}$), phase (or $\hat{Z}$), and bit-phase (or $\hat{Y}$) flip errors
result in a more rapid relaxation of the expected payoff
to the classical value,
as does depolarization,
with similar behaviour for these error types for $p < 0.5$.

The results for the bit, phase, and bit-phase flip errors can be understood as follows.
As the error probability is increased towards $\frac{1}{2}$ each qubit is reduced
to an equal superposition of the $|0\rangle$ and $|1\rangle$ states,
the optimal classical strategy,
and hence the classical payoff results.
Two phase flip errors per qubit will cancel each other out
so the curve for phase flip errors is symmetrical about $p=\frac{1}{2}$
since the errors are applied twice (see figure~\ref{f:qgame}).
For bit flip errors, the system approaches
an equal superposition of states with an even number of zeros and ones
as the error rate approaches one,
giving a zero payoff.
This effect dominates in the case of bit-phase flip error.
In these cases a new NE profile
where all agents play $\hat{M}(\pi/2, \pi/16, -\pi/16)$
emerges for $p > \frac{1}{2}$.
This profile yields the optimum (quantum) payoff.

\begin{figure}
\begin{center}
\includegraphics[width=9cm]{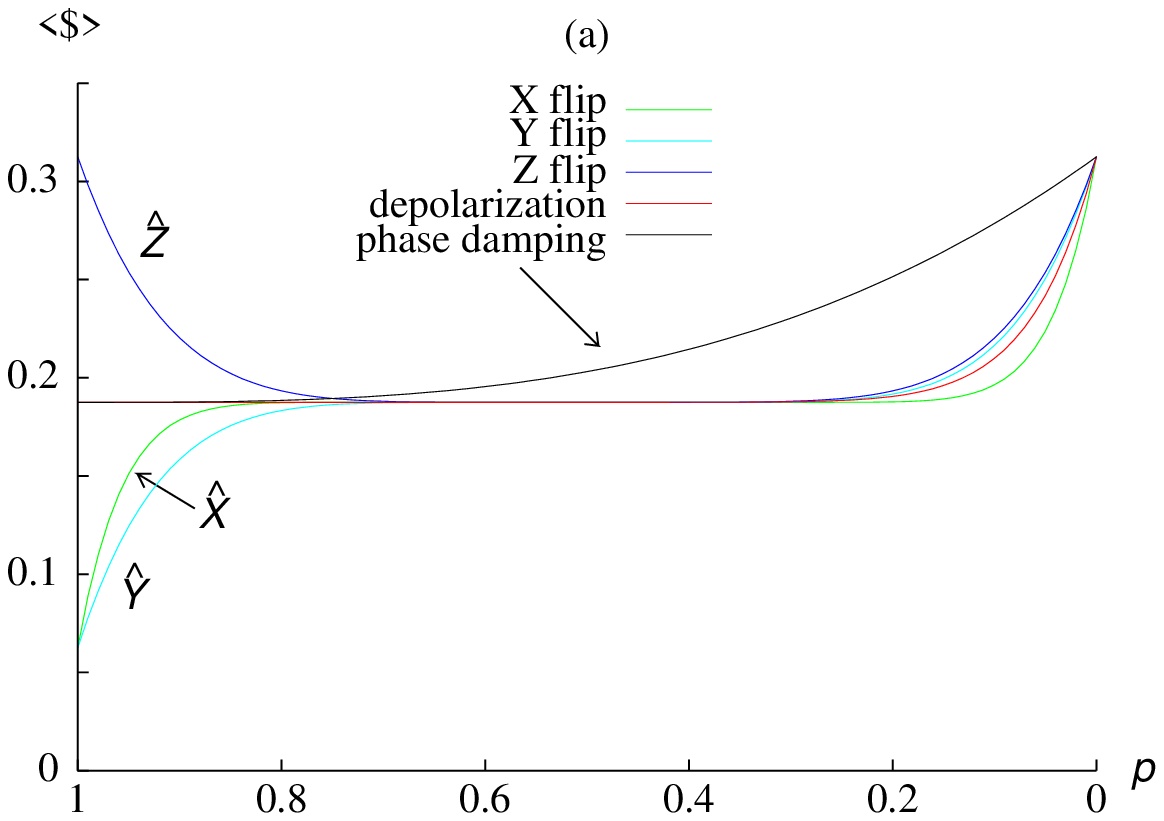}

\vspace*{1cm}

\includegraphics[width=9cm]{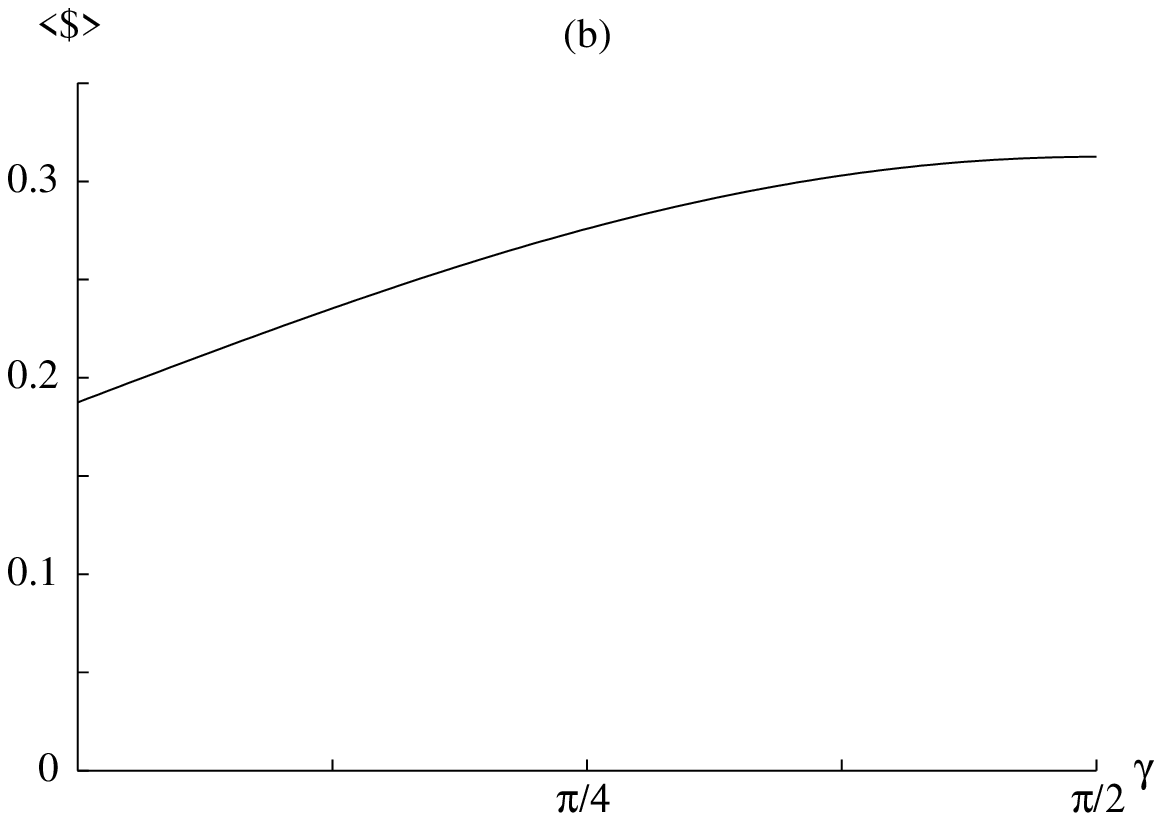}
\end{center}
\fcaption{\label{f:min6-decoh}(a) The Nash equilibrium payoff in an $N=6$ player quantum Minority game
as a function of the decoherence probability $p$.
The decoherence goes from the unperturbed quantum game at $p=0$ (right)
to maximum decoherence at $p=1$ (left)
The curves indicate decoherence by phase damping (black),
depolarization (red),
bit flip errors (green),
phase flip errors (blue)
and bit-phase flip errors (blue-green).
Compare this with (b) the Nash equilibrium payoff for $N=6$ as a function of the entangling
parameter $\gamma$ [\eref{J}].}
\end{figure}

\begin{figure}
\begin{center}
\includegraphics[width=9cm]{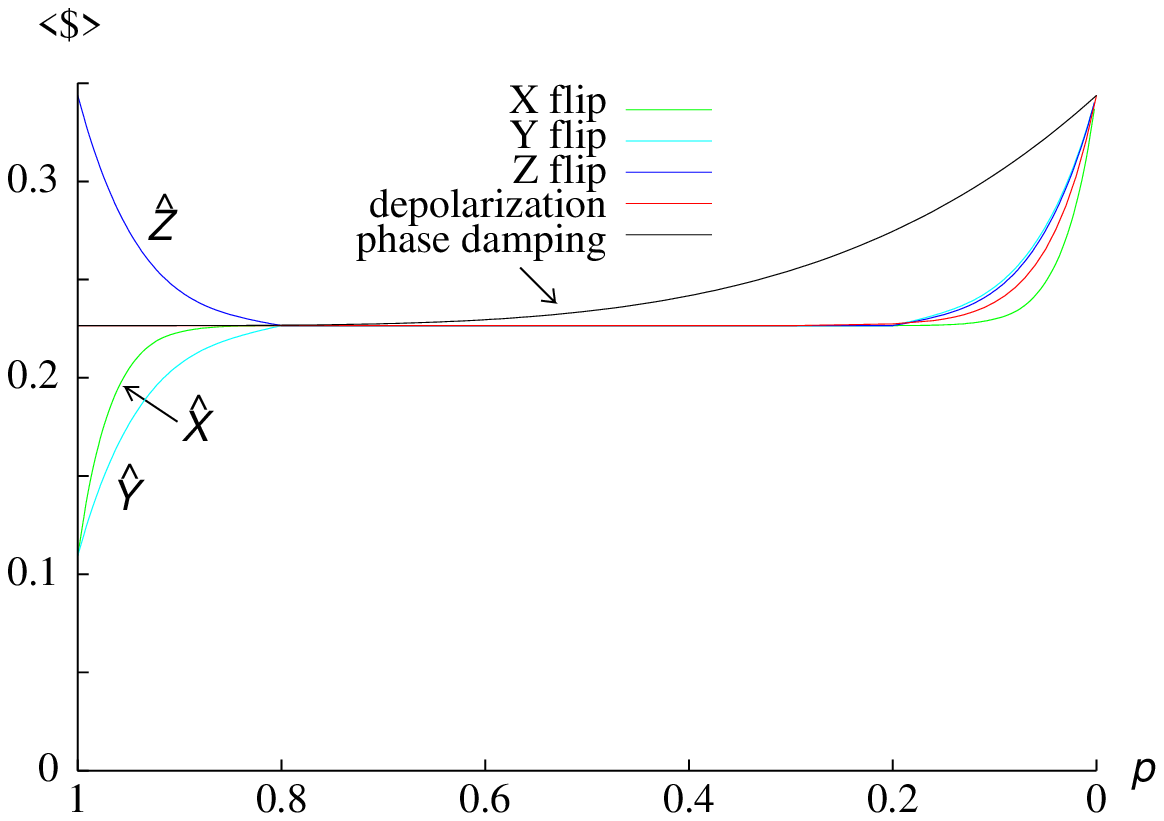}
\end{center}
\fcaption{\label{f:min8-decoh}The Nash equilibrium payoff in an $N=8$ player quantum Minority game
as a function of the decoherence probability $p$.
The decoherence goes from the unperturbed quantum game at $p=0$ (right)
to maximum decoherence at $p=1$ (left)
The curves indicate decoherence by phase damping (black),
depolarization (red),
bit flip errors (blue),
phase flip errors (green)
and bit-phase flip errors (blue-green).}
\end{figure}

Figures~\ref{f:min6-decoh} and \ref{f:min8-decoh} show the results
for the six and eight player quantum MG with decoherence.
Decoherence reduces the expected payoff for the NE strategy to the classical level
more quickly as $N$ increases
as a result of the increasing fragility of the GHZ state.

The entanglement that gives rise to the quantum enhancement in the payoff
is a global property of the $N$ qubits.
Decoherence in any qubit affects all the players equally,
not just the owner of the affected qubit.
In the case of phase damping---see \eref{measure}---where
$N_1$ players have decoherence probability $p_1$
and $N_2$ players have decoherence probability $p_2$,
the expected NE payoff for all the players can be expressed as
\begin{equation}
\langle \$ \rangle = \langle \$ \rangle_{\rm Cl} \:+\: (\langle \$ \rangle_{\rm Q} - \langle \$ \rangle_{\rm Cl})
	(1 - p_1)^{(N_1/2)} (1 - p_2)^{(N_2/2)}
\end{equation}
where the subscripts ${\rm \scriptstyle Cl}$ and ${\rm \scriptstyle Q}$ refer to classical and quantum, respectively.
Expressions for other types of errors are more complex
but remain equal for all players.
The fact that all the players score equally has relevance to the application of quantum error correction:
no player can advantage themselves over the other players by using quantum error correction on their qubit.
Instead, all players benefit equally.
The situation would be different, however,
if a subset of the players shared an entangled set of qubits
while the others were not entangled.

\section{CONCLUSION}
\label{sec-conc}
We have considered a quantum version of an $N$-player Minority game
where agents individually strive to select the minority alternative
out of two possibilities.
Entanglement amongst the qubits representing the players' selection
offers the possibility of enhancing the payoffs to the players
compared with the classical case
when the number of players is even.

When decoherence is added to the quantum Minority game,
the Nash equilibrium payoff is reduced as the decoherence is increased,
as one would expect.
However, the Nash equilibrium remains the same
provided the decoherence probability is less than $\frac{1}{2}$,
and is still the best result for the group
that can be achieved in the absence of cooperation.
The effect of depolarization, bit, phase, or bit-phase flip errors
reduces the expected payoff to the classical level
for an error probability of greater than approximately 0.2,
with the drop off being steeper as the number of players increases.
A more gradual reduction is seen with phase damping,
with an expected payoff above the classical level
unless the decoherence is maximized.

All players are equally disadvantaged by decoherence in one of the qubits.
Hence no player,
or group of players,
can gain an advantage over the remainder by utilizing quantum error correction
to reduce the error probability of their qubit.
However, the consideration of different forms of entanglement,
or partial-entanglement,
in the initial state is an interesting area for future study.
The simplicity and possible application of the Minority game suggests
that the study of the quantum version
is relevant to the theory of quantum information and quantum entanglement.

\nonumsection{Acknowledgments}
\noindent
Thanks go to Derek Abbott of The University of Adelaide
for reviewing an earlier draft of this manuscript,
to my colleagues Austin Fowler and Andy Greentree of Melbourne University,
to Charley Choe of Oxford University, and to Johannes Kofler of Universit{\"{a}}t Wien,
for their ideas and helpful discussion. 
Funding for AF was provided by the Australian Research Council grant number DP0559273.
LH is supported in part by the Australian Research Council, the Australian
government, the US National Security Agency,
the Advanced Research and Development Activity
and the US Army Research Office under contract number W911NF-04-1-0290.


\nonumsection{References}

\end{document}